\documentstyle[preprint,aps]{revtex}

\begin{document}
%\twocolumn
%\draft
%\today
%\documentstyle[preprint,revtex,eqsecnum]{aps}
%\flushbottom
\preprint{\vbox{
\hbox{IFT-P.032/95}
\hbox{IFUSP/P-1166}
%\hbox{DCR/TH pq-95}
%\hbox{hep-ph/9507264}
\hbox{July 1995}
}}
\title{A remark on the tau-neutrino mass limit}
\author{M.M. Guzzo\footnote{E-mail: guzzo@ifi.unicamp.br} }
\address{
 Instituto de F\'\i sica Gleb Wataghin \\
Universidade Estadual de Campinas, UNICAMP\\
13083-970 -- Campinas, SP\\
Brazil}
\author{ O.L.G. Peres\footnote{E-mail: orlando@vax.ift.unesp.br},
V. Pleitez\footnote{E-mail: vicente@vax.ift.unesp.br} }
\address{ Instituto de F\'\i sica Te\'orica\\
Universidade Estadual Paulista\\
Rua Pamplona, 145\\
01405-900 -- S\~ao Paulo, SP\\
Brazil}
\author{R. Zukanovich Funchal\footnote{E-mail: zukanov@charme.if.usp.br} }
\address{ Instituto de F\'\i sica da Universidade de S\~ao Paulo\\
05389-970 C.P. 66318 -- S\~ao Paulo, SP\\
Brazil}
\maketitle
\newpage
\begin{abstract}
We point out that the usual experimental upper bounds on
the ``tau-neutrino mass'' do not apply if neutrino mixing is considered.
The suppression of the population of the tau decay spectrum near
the end-point, caused
by mixing, may be compensated by an enhancement due to a resonant
mechanism of hadronization. It is necessary therefore to analyse
the whole spectrum to infer some limit to the ``tau-neutrino mass".
We argue that, consequently, neutrino mixing evades the objection
to interpret KARMEN anomaly as a heavy sequential neutrino.
\end{abstract}
%14.60.Lm ordinary neutrinos 14.60.Pq neutrino mass and mixing

In the near future, one of the most important issue
to be set by experimentalists, is the one of neutrino masses. Up to now, only
upper bounds have been established for the three type of
neutrinos~\cite{numass,lampf}.
In particular, concerning the ``tau-neutrino mass'', the upper
experimental limits for ``$m_{\nu_\tau}$'' are 31 MeV~\cite{nutau,pdg94} or
29 MeV~\cite{beijing}. Recent analysis performed by OPAL~\cite{opal} and
ALEPH~\cite{aleph} Collaborations, using
for the first time a two-dimensional technique i.e., invariant mass $\times$
total energy of the charged hadrons in the decay $\tau \to 5 \pi
\nu_\tau$ (OPAL) and $\tau\to 5\pi(\pi^0)\nu_\tau$ (ALEPH), give
``$m_{\nu_\tau}'' < 74 $ MeV and $<24$ MeV, respectively~\cite{fn}.
All these limits nevertheless come from analyses of the end point of
the hadronic invariant mass distribution  of the  $\tau$ decays
assuming no mixing effects among different neutrino species.
Usually, it is assumed, and in fact it seems reasonable, that
after the LEP measurements of the $Z$ width the existence
of sequential neutrinos with masses between the above upper bound
and $M_Z/2$ has become
unlikely. Nevertheless, this may not be the case if mixing does exist.

We will argue in the following that the presence of mixing among the
three generations can radically change our knowledge about the upper
bounds for the ``tau-neutrino mass''. Furthermore, a model independent
limit for this quantity is not possible if only the end point of the
spectrum is considered.

After the works of Shrock~\cite{rs}, it has been well known that massive
neutrinos will produce additional peaks determined by $m_i$
(mass of the $\nu_i$
neutrino, $i=1,2,3$) in the charged lepton spectrum in the decays
$\pi\to e(\mu)\nu_{e(\mu)}$~\cite{bri} or $K\to
e(\mu)\nu_{e(\mu)}$~\cite{kaons}.
Assuming Dirac neutrinos, the weak eigenstates $\nu_l$, $l=e,\mu,\tau$
are related to the mass eigenstate $\nu_i$  by $\nu_l=\sum_iV_{li}\nu_i$, where
$V$ is a $3\times 3$
unitary matrix.
The first processes are then a sensitive test of mixing
for neutrinos of mass in the range 50 to 130 MeV.
The no observance of peaks lead to the upper limit for the
mixing matrix elements  $\vert V_{ei}\vert^2<10^{-7}$~\cite{bri}.
However, in analyzing the data it was assumed that only one massive
neutrino couples to electrons, i.e.,
\begin{equation}
R_{ei}=\frac{\Gamma(\pi\to e \nu_i)}{\Gamma(\pi\to e\nu_1)}
=\vert V_{ei}\vert^2\rho(\delta_e,\delta_i),
\label{pi1}
\end{equation}
with $\nu_1$ being a massless neutrino and $\rho(\delta_e,\delta_i)$ is a
kinematic factor with $\delta_i=(m_{i}/m_\pi)^2$ and  $\delta_e=m^2_e/m^2_\pi$.
However, the pion decay constraint does not apply if the neutrinos
have masses about or higher than pion mass $\sim140\,\mbox{MeV}/c^2$.
Similar analyses have been done in kaon decays
but they are not  statistically significant~\cite{rev}. The difference between
two and three generation treatments (pointed out recently~\cite{tau4,nos})
implies
that the constraints above cannot be used straightforward for the three
generation case. That is, it is possible that there exist truly three neutrino
mixing effects which are not reduced to two neutrino mixing effective ones.
So, it is important to reanalize the data on pion and kaon decays from the
point
of view of three generations. We will return to this point at the end of this
paper.

With respect to the ``tau-neutrino mass" measurement in the decay
$\tau\to 5\pi\nu_\tau$ we would like to emphasize
the following. Assuming also a three generation mixing and that $\nu_\tau$ in
that decay is
a mixing of mass eigenstates being $\nu_3$ heavier than $\nu_{1,2}$ we get
\begin{equation}
\frac{d \Gamma}{d q^2} = \frac{G_\mu^2 V^2_{KM}}{8 \pi m^3_\tau
(\sin^2\beta\sin^2\gamma+\cos^2\gamma)\cos^2\beta} \,
\left[ F_0 +
 c^2_\gamma c^2_\beta (F_3 - F_0)\right ] h(q^2) ,
\label{gamma}
\end{equation}
where
\begin{equation}
F_0 = \omega(q^2,m^2_\tau,0)\lambda^{\frac{1}{2}}(m^2_\tau,q^2,0),
\quad
F_3 = \omega(q^2,m^2_\tau,m^2_3)
\lambda^{\frac{1}{2}}(m^2_\tau,q^2,m^2_3),
\label{F3}
\end{equation}
and
\begin{equation}
\omega (q^2,m^2_\tau,m^2_3) = (m^2_\tau -q^2)(m^2_\tau +2 q^2) -
m^2_3 (2 m^2_\tau -q^2 - m^2_3),
\label{omega}
\end{equation}
where $\lambda$ is the usual triangular function, $q^2$ is the
transferred momentum,  $h(q^2)$ contains the hadronic structure and
$V_{KM}^2$ denotes the quark mixing angles, in this case
$V_{KM}^2=V_{ud}^2$. We have used the parametrization of the mixing
matrix of Ref.~\cite{tau4}.

With the values of the mixing angles of Ref.~\cite{tau4}, it is possible
to see from Eqs.~(\ref{gamma})-(\ref{omega})
that the resulting spectra when we employ
nonvanishing value for $m_3$ are systematically about 20 times
smaller than the spectrum obtained in the massless case, for regions
with values of the hadronic invariant mass after the kink,
$m_{h}= \sqrt{q^2}>m_\tau-m_3 $ i.e., near the end point of the spectrum.
This can be better appreciated in Fig. 2 of Ref.~\cite{tau4}.

Experimental results show events with hadronic invariant mass very
close to the charged tau lepton mass $m_\tau$. Since there exists the
mentioned suppression of a factor 20 for spectra with $m_3 \neq 0$,
this experimental fact might be interpreted as an absence of kinks
or, even if there is a kink, it is localized so close to $m_\tau$
that  $m_3$ is constrained as discussed above~\cite{pdg94,beijing,opal}.

Nevertheless such kind of analysis takes into account only the phase
space contribution to the $\tau \to 5 \pi \nu_\tau$ decay spectrum. One
has to consider also the effect of possible resonant intermediate
channels to this decay. This can radically change the population of
this spectrum near the end point. In fact in Ref.~\cite{concha} some
channels as $\tau \to \rho^0 \rho^0 \pi\nu_\tau$ were considered.  They
found an enhancement of the population of the spectrum near the
massless end point of a factor of 30 when compared with the pure phase
space contribution. The ALEPH Collaboration has recently performed a
similar analysis and has also demonstrated that some resonant channels
will equally increase the population of the spectrum in that
region~\cite{aleph1}.

The spectral function $h(q^2)$ appearing in Eq. (\ref{gamma})
does not depend on
mixing parameters as well as the $m_3$ mass. Therefore this same
function will alter
the spectra where mixing is implemented and $m_3$ assumes nonvanishing
values.
In Fig. 1 we give an example of how this function, calculated assuming
 the resonant intermediate state $\rho^0 \rho^0 \pi$, significantly
 change the spectrum near the end point and can compensate the phase
space suppression due to the mixing in this region.  Based on Ref.
\cite{tau4}, we use $m_3=165 $ MeV, and $11.54^o <\beta<12.82^o$ and
$\gamma < 4.05^o$ to illustrate this phenomenon. (Note also that
Bottino
et al., in a recent analysis~\cite{ita}, have also found $m_3$
larger than the usually accepted experimental value.)

 From this figure we see that the simple investigation of the
end point part of the spectrum may not be sufficient to infer any
information about the ``tau-neutrino mass".
Both possibilities, the massless spectrum and that
one obtained when a heavy neutrino is taken into account, overlap in
that region, making it impossible to distinguish between them unless the
contributions to this decay from resonant channels is well understood.
This is not the case at present.

Furthermore the hadronization process which has to be taken into
account in the calculation of intermediate channel contributions to the
$\tau$-decay spectrum is very model dependent. We can affirm that a
conclusive and model independent limit for the ``tau-neutrino mass''in
the hadronic decays is not
possible if we have experimental information only near the end point of
the spectrum.

 Current limits for the ``tau-neutrino mass" do not apply if neutrino
mixing is
considered, even a small mixing angle in vacuum may be  sufficient to
allow for the
mentioned enhancement due to hadronization.

Let us go back now to the pion decays. Recently KARMEN Collaboration has
reported,
based on the charged and neutral current reaction $^{12}C(\nu_e,e)^{12}N$
and $^{12}C(\nu,\nu')^{12}C$, that a new neutral particle $x$, produced in the
$\pi^+\to \mu^+x$, must has a mass of $m_x=33.9$ MeV~\cite{karmen}.
Identification of this massive neutral particle with the $\nu_3$ has been
ruled out by the ALEPH upper mass limit bound~\cite{aleph,karmen}.
The possibility of the $x$-particle  being a mainly-sterile neutrino has been
pointed out in Ref.~\cite{varger}.
However, once one has convinced ourselves that the current upper values
for the ``tau-neutrino mass" are not valid when we have mixing,
the possibility arises that KARMEN's $x$ could still be interpreted as
a heavy sequential neutrino.

The main point of this letter is that, to investigate if a large $m_3$
mass hypothesis is realized or not by nature, experimentalists must
measure and fit the shape of the whole spectrum of  $\tau \to 5 \pi
\nu_\tau$ decay, including regions far from the end point. Indications
of a nonvanishing $m_3$ would be given by the characteristic kink
vertically displaced from the phase space contribution to the spectrum
by the spectral function.  Although the form of this function is very
model depend, the projection of the kink on the hadronic mass axis
would be, in principle, model independent and a good observable for
experimentalists.

Since the branching ration for
this decay is of the order of $10^{-4}$, present data are not sufficient
 to build the full relevant spectrum looking for kinks.  This analysis
requires large
statistics. Tau factories would be necessary to achieve this
goal.

\newpage
\acknowledgements

We thank Funda\c c\~ao de Amparo \`a Pesquisa do Estado de
S\~ao Paulo (FAPESP) for full financial support (O.L.G.P.) and
Con\-se\-lho Na\-cio\-nal de De\-sen\-vol\-vi\-men\-to Cien\-t\'\i
\-fi\-co e Tec\-no\-l\'o\-gi\-co (CNPq) for partial financial support
(M.M.G., V.P. and R.Z.F.).

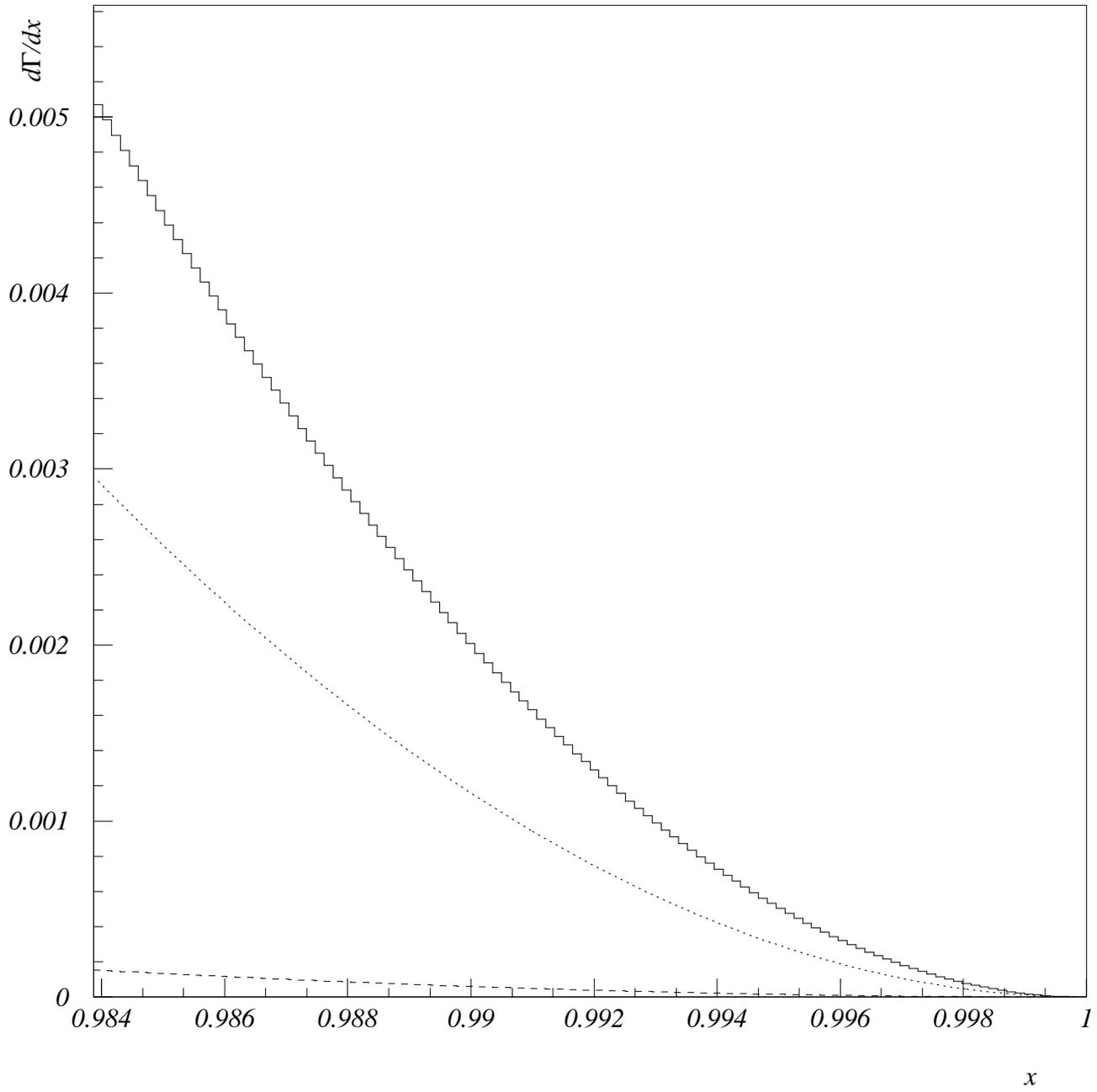
\end{document}